\documentclass[twocolumn,aps,prl,superscriptaddress,showpacs]{revtex4}
\usepackage{amssymb}
\usepackage{graphicx}
\usepackage{amsmath}
\usepackage{amsfonts}
\usepackage{color}

\begin{document}


\title{Chern Kondo Insulator in an Optical Lattice}

\author{Hua Chen}
\affiliation{International Center for Quantum Materials and School of Physics, Peking University, Beijing 100871, China}
\affiliation{Collaborative Innovation Center of Quantum Matter, Beijing 100871, China}
\author{Xiong-Jun Liu\footnote{Materials and Correspondence should be addressed to: xiongjunliu@pku.edu.cn}}
\affiliation{International Center for Quantum Materials and School of Physics, Peking University, Beijing 100871, China}
\affiliation{Collaborative Innovation Center of Quantum Matter, Beijing 100871, China}
\author{X. C. Xie}
\affiliation{International Center for Quantum Materials and School of Physics, Peking University, Beijing 100871, China}
\affiliation{Collaborative Innovation Center of Quantum Matter, Beijing 100871, China}

\date{\today}

\begin{abstract}
We propose to realize and observe Chern Kondo insulators in an optical superlattice with laser-assisted $s$ and $p$ orbital hybridization and synthetic gauge field, which can be engineered based on the recent cold atom experiments. Considering a double-well square optical lattice, the localized $s$ orbitals are decoupled from itinerant $p$ bands and are driven into a Mott insulator due to strong Hubbard interaction. Raman laser beams are then applied to induce tunnelings between $s$ and $p$ orbitals, and generate a staggered flux simultaneously. Due to the strong Hubbard interaction of $s$ orbital states, we predict the existence of a critical Raman laser-assisted coupling, beyond which the Kondo screening is achieved and then a fully gapped Chern Kondo phase emerges, with the topology characterized by integer Chern numbers. Being a strongly correlated topological state, the Chern Kondo phase is different from the single-particle quantum anomalous Hall state, and can be identified by measuring the band topology and double occupancy of $s$ orbitals. The experimental realization and detection of the predicted Chern Kondo insulator are also proposed.
\end{abstract}
\pacs{}

\maketitle

{\it Introduction}.-- The search for new quantum states with nontrivial topology has become an exciting pursuit in condensed matter physics. The recent important examples are the theoretical prediction and experimental discovery of the 2D and 3D topological insulators (TIs)~\cite{hasan10,qi11}. Recently, a few heavy-fermion materials (e.g. $\text{SmB}_6$) were predicted to be candidates of time-reversal-invariant topological Kondo insulator (TKI), which originates from the hybridization between itinerant conduction bands and strong correlated $f$ electrons~\cite{Dzero10,Alexandrov13,lu13}.
It was shown that screening the local moments by conduction electrons leads to an insulating gap
in the Kondo regime.
The proposed scenario of TKIs is consistent with the transport measurement~\cite{wolgast13,Kim14,Luo15}, angle-resolved photoemission spectroscopy~\cite{jiang13,Neupane13,Xu14}
and scanning tunneling spectroscopy~\cite{Ruan14,Roessler14}. Note that the essential difference between a TKI and conventional TIs is not topological classification, but the strong correlation physics for $f$ electrons which are absent in conventional TIs. To measure such strong correlation physics can directly distinguish a TKI from conventional TIs, while this might be a challenging task for condensed matter systems.

In this letter, we propose to realize and observe a strongly correlated quantum anomalous Hall (QAH) phase, called Chern Kondo (CK) insulator, in an optical lattice, motivated by the recent rapidly developing new technologies for cold atoms. These technologies include the Raman coupling scheme~\cite{Ruseckas2005,Osterloh2005,Liu2009} used to create spin-orbit interactions~\cite{Lin,Pan,MIT,Wang}, laser-assisted tunneling or shaken lattices to generate synthetic gauge fields~\cite{Bloch2011,Bloch2013,Ketterle2013a,Struck2013}, and optical control of Feshbach resonance~\cite{Walraven1996,Lett2000,Denschlag2004,Takahashi2008,Bauer2009,Ye2011,ChengChin2015}. Compared with solid state systems, the cold atoms can offer extremely clean platforms with full controllability to study many-body physics and topological phases~\cite{Chuanwei2008,Sato2009,Liu10,Goldman2010,Zhu2011,Liu2014,Goldman2014,Zhai2015}. Here, we consider a double-well square lattice with Raman-coupling-assisted $s$-$p$ orbital hybridization to observe CK insulating phases. Due to the strong Hubbard interaction of $s$ orbital states, the Kondo screening is achieved when the applied Raman coupling exceeds a critical value, and then a nonzero renormalized $s$-$p$ orbital hybridization drives the system into a fully gapped CK phase. The parent state of $s$ orbital is a Mott insulator which has no double occupancy due to the Hubbard gap, and this property keeps in the CK regime, signifying an essential difference from single-particle QAH insulators. With this unique feature the CK insulators can be identified by measuring band topology and double occupancy of $s$ orbital in experiment.

\begin{figure}
	\includegraphics[width=0.48\textwidth]{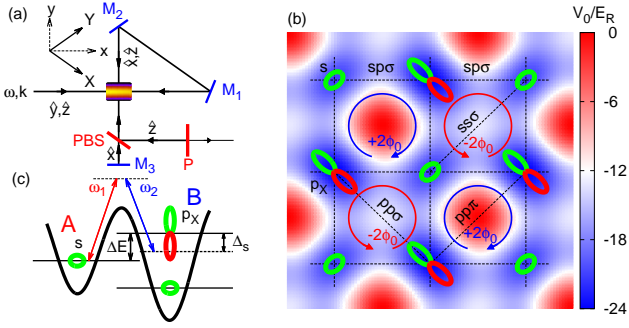}
	\caption{(color online) (a) Sketch of generating double-well checkerboard superlattice by an incident laser beam with in-plane ($\hat y$) and out-of-plane ($\hat z$) polarized components. The $\hat z$-component is partially reflected by the mirror $P$. This configuration generates a 2D checkerboard lattice $V_\text{Latt}(x,y) = -V_0 [\cos^2(k x) + \cos^2(k y)]-V_1 \sin^2[\frac{k}{2}(x-y)]-V_2 \cos^2[\frac{k}{2}(x+y)]$~\cite{SI}. (b) The lattice potential profile for $\{V_0,V_1,V_2\}=\{8,8,5\}E_R$, with $E_R$ being the recoil energy. (c) Two Raman lasers are applied to induce $s$-$p_X$ orbital hybridization and create staggered flux pattern in (b).}
	\label{fig:OL}
\end{figure}
{\it Model}.--For the realization we introduce a novel bipartite square optical lattice created with the setup in Fig.~\ref{fig:OL}(a), which has a staggered energy difference between the neighboring A and B sites. This double-well checkerboard lattice can be realized based on NIST experiments~\cite{sebby06,SI}, and more details will be presented later. The minimal model for realizing the CK insulators includes several basic ingredients. First, the sublattices A and B are anisotropic along the local coordinates $\hat{X}/\hat{Y}=\pm \hat{x}+\hat{y}$.
The relevant orbitals for our consideration are the $s$ orbital at A sites and $p_X$ orbital at B sites [Fig.~\ref{fig:OL}(b)], with the intraorbital AA/BB hopping couplings along the diagonal directions.
In the moderate to deep lattice regime we can verify that $t_{ss\sigma}$ is much less than $t_{pp\sigma}$, leading to a relatively flat band for s-orbital. Accordingly, in such regime the transverse tunneling $t_p^Y$ between $p_X$-orbitals is much weaker than axial tunneling $t_p^X$~\cite{lsacsson05,liu06}. Thus for the practical consideration, we neglect the $\pi$-bond hoppings of the $s$ and $p_X$ orbitals by setting $t_p^Y=t_s^X=0$, and for the $\sigma$-bond hopping we have $|t_p^X|\gg |t_s^Y|$.
Secondly, similar as the recent experimental studies~\cite{Bloch2011,Bloch2013,Ketterle2013a}, we consider an onsite energy difference $\Delta E=E_B-E_A$ to suppresses the bare tunneling couplings between $s$ and $p_X$ orbitals. On the other hand, the neighboring $s$-$p_X$ hybridization can be induced by a standard two-photon Raman coupling in the following form $V_R=V_m\cos(\delta\omega t+k_Ry)$,
where $V_m$ is the coupling amplitude, $\delta\omega$ and $\bold k_R=k_R\hat y$ represent the differences of frequencies and wave vectors between two Raman photons, respectively [Fig.~\ref{fig:OL}(c)]~\cite{Bloch2011,Bloch2013,Ketterle2013a}. The Raman potential drives the neighboring hopping between $s$ and $p_X$ orbitals when $\delta\omega$ compensates their energy difference $\Delta E$. Moreover, this laser-assisted hopping along $y$ direction is associated with a phase $\phi_0=k_Ra$ $(-k_Ra)$ if the hopping is toward (away from) B sites with A being lattice constant, rendering a staggered flux pattern depicted in Fig.~\ref{fig:OL} (b) with the flux $\Phi=2\phi_0$.
Finally, we consider a (pseudo)spin-$1/2$ two-copy version of the $s$-$p_X$ band model, so the Hubbard interactions can be introduced for the spinful fermions (e.g. $^6$Li or $^{40}\text{K}$) and denoted by $U_s$ ($U_p$) for the $s$ ($p_X$) orbitals. With these ingredients we can write down the effective Hamiltonian by $H=H_0+H_{\rm int}$, where
\begin{eqnarray}
H_0 &=& \sum_{i\sigma} [
	-t_s^Y s^\dagger_{i\sigma}s_{i\pm\hat{Y}\sigma}-\Delta_s s^\dagger_{i\sigma}s_{i\sigma}
	+t_p^X p^\dagger_{Xi\sigma}p_{Xi\pm\hat{X}\sigma}]\nonumber \\
	&+& \sum_{\left<ij\right>\sigma} [F(\mathbf{r})s_{i\sigma}^\dagger p_{Xj\sigma}\delta_{j,i+\mathbf{r}}+\text{h.c.}]\label{eq:tb}\\
H_\text{int}&=& U_s\sum_i \hat{n}_{si\uparrow}\hat{n}_{si\downarrow}
	+U_p\sum_j \hat{n}_{p_Xj\uparrow}\hat{n}_{p_Xj\downarrow}.
\end{eqnarray}
Here $\Delta_s=\Delta E-\delta\omega$ is the two-photon detuning for the Raman coupling [Fig.~\ref{fig:OL}(c)], $s_{i\sigma}$ and $p_{Xj\sigma}$ ($s^\dagger_{i\sigma}$ and $p^\dagger_{Xj\sigma}$) are annihilation (creation) operators of $s$ and $p_X$ orbitals, respectively, $\hat{n}_{si\sigma}=s^\dagger_{i\sigma}s_{i\sigma}$ and $\hat{n}_{p_Xj\sigma}=p^\dagger_{Xj\sigma}p_{Xj\sigma}$
The Raman coupling induced hybridization satisfies $F(\mathbf{r})=\pm t_{sp}e^{i\phi_0}$ for $\mathbf{r}=\pm a\hat y$
and $F(\mathbf{r})=\pm t_{sp}$ for $\mathbf{r}=\pm a\hat x$, with $t_{sp}$ being the induced hybridization strength.

In the single particle regime the present setup can realize QAH effect~\cite{Liu10}. We rewrite $H_0$ in the $\bold k$ space $H_0=\sum_{\mathbf{k}\sigma}\mathcal{C}^
\dagger_{\mathbf{k}\sigma}\mathcal{H}_0(\mathbf{k})\mathcal{C}_{\mathbf{k}\sigma}$
with $\mathcal{C}_{\mathbf{k}\sigma}=(s_{\mathbf{k}\sigma},p_{X\mathbf{k}\sigma})^T$.
The Bloch Hamiltonian takes the form
\begin{eqnarray}
\mathcal{H}_0(\mathbf{k})=
d_0(\mathbf{k}) \tau_0
+d_x(\mathbf{k}) \tau_x
+d_y(\mathbf{k}) \tau_y
+d_z(\mathbf{k}) \tau_z,
\label{eq:bloch}
\end{eqnarray}
where $d_x(\mathbf{k})=2t_{sp}\sin k_y\sin\phi_0, d_y(\mathbf{k})=2t_{sp}\sin k_x+2t_{sp}\sin k_y\cos\phi_0$,  $d_{0/z}(\mathbf{k})=\pm t_p^X\cos k_X-t_s^Y\cos k_Y-\Delta_s/2$,
and the Pauli matrices $\mathbf{\tau}_{x,y,z}$ act on orbital space.
The single-particle spectra, given by $E^{\pm}_{\mathbf{k}}=d_0(\mathbf{k})\pm \sqrt{d_x(\mathbf{k})^2+d_y(\mathbf{k})^2+d_z(\mathbf{k})^2}$, has node points if $\phi_0=m\pi$ with $m\in\mathbb{Z}$, and is gapped when $\phi\neq m\pi$ and $\Delta_s\neq\pm2(t_p^X+t_s^Y)$. The gapped single-particle system is in the QAH phase if $|\Delta_s|<2(t_p^X+t_s^Y)$~\cite{Liu10}. On the other hand, as studied below, a positive detuning $\Delta_s>0$ will be applied to reach the CK phase.

{\it Strongly interacting regime and Kondo transition}.--To study the Kondo phase, we consider the strongly interacting regime for $s$-orbital states with $U_s/t_s^Y\gg1$, and a weak interaction for $p_X$ orbital states, satisfying $U_s/t_s^Y \gg U_p/t_p^X$.
To treat the interaction appropriately, the weak $U_p$ terms are decoupled by mean-field approximation
$\hat{n}_{p_Xj\uparrow}\hat{n}_{p_Xj\downarrow} \simeq \sum_{\sigma}n_{p_Xj\sigma}\hat{n}_{p_Xj\bar{\sigma}}
	-n_{p_Xj\uparrow}n_{p_Xj\downarrow}$.
In the strong interacting limit $U_s/t_s^Y\gg1$,
the state of double-occupation $^{2}s$ on $s$ orbital is prohibited.
The system can be studied by the slave boson mean field theory
with vanishing doublon~\cite{Kotliar86}.
We introduce the slave-boson operator $b_i^\dagger$($b_i$) at $i$-th site
to describe the
creation(annihilation) holon state $^{0}{s}$.
Moreover, we consider $Z_\sigma s_{i\sigma}^{\dagger} p_{Xj\sigma}$
instead of $s_{i\sigma}^{\dagger}p_{Xj\sigma}$ in eq.~(\ref{eq:tb})
with $Z_\sigma = \frac{\bar{b}\bar{s}_\sigma}
{\sqrt{1-\bar{s}_\sigma^2}\sqrt{1-\bar{b}^2-
\bar{s}_{\bar{\sigma}}^2}}$.
Here the slave operators in uniform mean field approximation
are replaced by time- and site- independent numbers, $\bar{s}_\sigma$ and $\bar{b}$.
The hopping $t_s^Y$ is also renormalized
by a factor of $Z_{\sigma}^2$, which preserves Luttinger volume
and captures the correct band renormalization.
To eliminate the redundant many-particle configurations of $s$ orbital,
the constraints
$
\lambda\left(
\sum_{\sigma}\bar{s}_{\sigma}^2+\bar{b}^2-1
\right)
$
and
$\sum_{i\sigma}\lambda_{\sigma}\left(
\hat{n}_{si\sigma}-\bar{s}_{\sigma}^2
\right)$
with $\lambda,\lambda_{\sigma}$ being the Lagrange multipliers are introduced.
The functional for free energy per site takes the form~\cite{SI}
\begin{eqnarray}
\mathcal{F}&=&-\frac{1}{N\beta}\sum_{\mathbf{k}\sigma,\pm}\ln \left(1+\exp[-\beta E^{\pm}_{\mathbf{k}\sigma}]\right)\nonumber\\
&&+\lambda\left(\sum_{\sigma}\bar{s}_{\sigma}^2+\bar{b}^2-1\right)
	-\sum_{\sigma}\lambda_{\sigma}\bar{s}_{\sigma}^2,
\end{eqnarray}
where $\beta^{-1}=k_\text{B}T$, the quasiparticle spectra of the lower $(-)$ and upper $(+)$ hybridized bands are given by $E^{\pm}_{{\bf k}\sigma}=\bigr[
\varepsilon_{p_X\mathbf{k}\sigma}+\varepsilon_{s{\bf k}\sigma}
\pm \sqrt{(\varepsilon_{p_X\mathbf{k}\sigma}-\varepsilon_{s{\bf k}\sigma})^{2}+4|Z_\sigma\mathcal{V}_\mathbf{k}|^2}\bigr]/2$, respectively, with
$\varepsilon_{p_X{\bf k}\sigma}=2t_p^X\cos k_X
+U_pn_{p_X\bar{\sigma}}$,
$\varepsilon_{s\mathbf{k}\sigma}=-2Z_\sigma^2 t_s^Y\cos k_Y-\Delta_s+\lambda_\sigma$
and $\mathcal{V}_\mathbf{k}=2it_{sp}\left(\sin\frac{k_X+k_Y}{2}
\exp\left[i\phi_0\right]+\sin\frac{k_X-k_Y}{2}\right)$.
The minimization of the free energy with respect to
the mean field parameters are determined by
$\frac{\partial \mathcal{F}}{\partial \xi}=0$ with $\xi=\bar{s}_\sigma,\bar{b},\lambda_\sigma,\lambda$,
which yields several coupled saddle-point equations
\begin{widetext}
	\begin{eqnarray}\label{eqn:selfconsistent}
	\{\lambda-\lambda_{\sigma^\prime},\lambda\}
	&=& \sum_{\sigma}\{
	\frac{Z_\sigma}{\bar{s}_{\sigma^\prime}}\frac{\partial Z_\sigma}{\partial \bar{s}_{\sigma^\prime}},
	\frac{Z_\sigma}{\bar{b}}\frac{\partial Z_\sigma}{\partial \bar{b}}\}
	\times
	\frac{1}{N}\sum_{{\bf k},\pm}\left[
	t_s^Y\cos k_Y\mp \frac{(\varepsilon_{p_X\mathbf{k}\sigma}-\varepsilon_{s\mathbf{k}\sigma})
		t_s^Y\cos k_Y+|\mathcal{V}_\mathbf{k}|^2}
	{\sqrt{(\varepsilon_{p_X\mathbf{k}\sigma}-\varepsilon_{s{\bf k}\sigma})^{2}+4|Z_\sigma\mathcal{V}_\mathbf{k}|^2}}\right]
	f(E_{{\bf k}\sigma}^{\pm})	\nonumber\\	
	\bar{s}_\sigma^2 &=& \frac{1}{2N}\sum_{{\bf k},\pm}
	\left[1\mp\frac{\varepsilon_{p_X\mathbf{k}\sigma}
		-\varepsilon_{s{\bf k}\sigma}}{\sqrt{(\varepsilon_{p_X\mathbf{k}\sigma}
			-\varepsilon_{s{\bf k}\sigma})^{2}+4|Z_\sigma\mathcal{V}_\mathbf{k}|^2}}\right]
	f(E^{\pm}_{{\bf k}\sigma}), \	\ \bar{b}^2= 1-\sum_\sigma \bar{s}_\sigma^2,
	\end{eqnarray}
\end{widetext}
where $f(E)=1/[e^{\beta(E-\mu)}+1]$ is the Fermi-Dirac distribution function with $\mu$ the chemical potential.
The half-filling condition for the entire system is fixed by the equation:
$1=\sum_{{\bf k}\sigma} \left[
f(E^{+}_{{\bf k}\sigma})+f(E^{-}_{{\bf k}\sigma})\right]/2N$, which sets the chemical potential
in the self-consistent calculation.

\begin{figure}
	\includegraphics[width=0.48\textwidth]{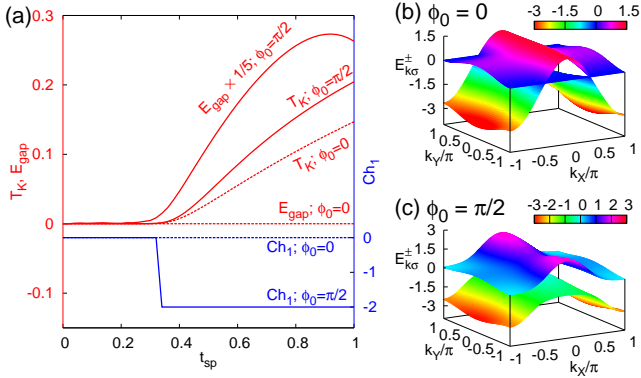}
	\caption{(color online) (a) Kondo temperature $T_K=\bar{b}^2$, direct band gap $\Delta$ and Chern number $\text{Ch}_1$
		as a function of hopping integral $t_{sp}$ with $(t_p^X,t_s^Y,\Delta_s,U_p)=(1,0.1,2,2)$ in unit of $t_p^X$ for gauge field $\phi_0=0$ and $\frac{\pi}{2}$. Band dispersion $E_{\mathbf{k}\sigma}^\pm$ with $t_{sp}=0.8t_{p}^X$, $\phi_0=0$ (b) and $\phi_0=\frac{\pi}{2}$ (c). Each band has two fold degeneracy with respect of spin projection $\sigma$.}
	\label{fig:tsp}
\end{figure}
The Kondo transition can be studied in the regime with $\Delta_s>0$, which implies that the onsite energy of the $s$-orbital is lower than the Fermi energy of the half-filled $p_X$-band. The direct hopping from $p_X$ orbital to $s$ orbital is however forbidden due to the Hubbard gap in the $s$ states, while the process of ``cotunneling" which is a second-order virtual process can occur between $s$ and $p_X$ orbitals. This process is known as the Kondo coupling. In the weak Raman coupling regime, the $s$-orbital states stay in a Mott insulator decoupled from the itinerant $p_X$ band, since the spectral weight of $s$-band vanishes at the Fermi energy. We thus expect that the emergence of the mean-field effective hybridization, namely, the Kondo coupling between $s$ and $p_X$ orbitals requires that the laser-assisted tunneling $t_{sp}$ exceeds a critical value, beyond which the Kondo screening is achieved and a fully gapped CK phase emerges.

We show the numerical results in Fig.~\ref{fig:tsp} by solving the mean-field equations~\eqref{eqn:selfconsistent} self consistently. We can see that the effective hybridization between $s$ and $p_X$ orbitals is strongly renormalized by a factor $Z_\sigma$ due to the Hubbard interaction in the $s$ orbital states. A critical laser-assisted tunneling $t_{sp}^c$ is observed and when $t_{sp}<t^c_{sp}$, the effective $s$-$p_X$ hybridization renormalizes to zero [see the red curves in Fig.~\ref{fig:tsp}(a)]. This is in sharp contrast to a single-particle system. Accordingly, increasing $t_{sp}$ to exceed $t^c_{sp}$, the coherent hybridization is gradually developed, leading to the formation of heavy quasi-particle bands as shown in Fig.~\ref{fig:tsp} (b) and (c). The local moments are screened by spin flipping through the cotunneling process. A direct bulk gap $E_{\rm gap}>0$ opens between the heavy quasi-particle bands for $\phi_0=\pi/2$. In contrary, if the gauge flux $\phi=0$, the $d_x(\mathbf{k})$ in eq.~(\ref{eq:bloch}) vanishes and the bulk has gapless nodal points.

\begin{figure}
	\includegraphics[width=0.48\textwidth]{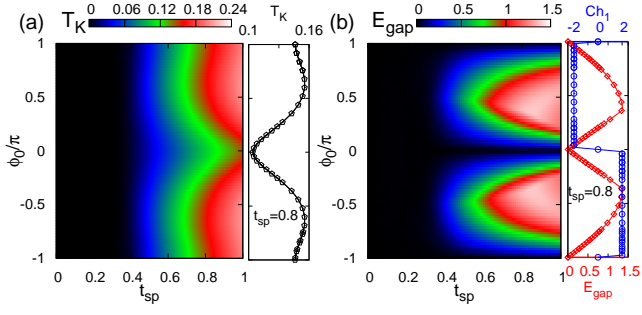}
	\caption{(color online) Main panel: (a) Image map of Kondo temperature $T_K\propto \bar{b}^2$ and (b) direct band gap $\Delta$ as a function of hopping integral $t_{sp}$ and gauge field $\phi_0$ with $(t_p^X,t_s^Y,\Delta_s,U_p)=(1,0.1,2,2)$. Subfigue: (a) Kondo temperature $T_K$(open pentagon) and (b) band gap $\Delta$(open square), Chern number $\text{Ch}_1$(open circle) versus $\phi_0$ at $t_{sp}=0.8t_p^X$.}
	\label{fig:phi}
\end{figure}

To analyze the topology of the bands, the Chern number for the $n$-th band is explicitly calculated by
\begin{eqnarray}
\text{Ch}_1&=& \int_\text{BZ} d^2\mathbf{k}
\sum_{m(\neq n) , \sigma}[f\left(E^n_{\mathbf{k}\sigma}\right)-f\left(E^m_{\mathbf{k}\sigma}\right)]\nonumber \\
	&\times&
	\frac{1}{2\pi}\frac{\text{Im}\left[
	\langle n\mathbf{k}\sigma\vert \hat{v}_X \vert m\mathbf{k}\sigma\rangle
	\langle	m\mathbf{k}\sigma\vert \hat{v}_Y \vert n\mathbf{k}\sigma\rangle\right]
}{(E^n_{\mathbf{k}\sigma}-E^m_{\mathbf{k}\sigma})^2},
\end{eqnarray}
where $\hat{v}_{X/Y}=\partial \mathcal{H}_\text{T}/\partial k_{X/Y}$ are velocity operators along $\hat{X}/\hat{Y}$ directions.
Fig.~\ref{fig:tsp}(a) shows nonzero Chern number for $\phi_0=\pi/2$. The nontrivial topological property originates from the band inversion between the $s$ and $p_X$ bands. The phase diagrams versus gauge field $\phi$ and $t_{sp}$ are shown in Fig.~\ref{fig:phi} (a,b), with the subfigure showing the gauge field dependence of $T_K$ and Chern number at a line cut of $t_{sp}=0.8t_p^X$. The $T_K$ and the direct gap $E_{\rm gap}$ show symmetric symmetric behavior with respect to $\phi_0=0$, while Chern number $\text{Ch}_1$ changes sign across this point.

{\it Experimental realization and detection}.--Now we turn to the model realization and detection of the CK insulating phase. The square superlattice potential $V_\text{Latt}(x,y)$ can be generated through the setup shown in Fig.~\ref{fig:OL} (a) with an incident laser field $\bold E=(E_1\hat y+E_2\hat z)e^{ik_0x}$~\cite{SI}. Neglecting all the irrelevant phase factors, the in-plane polarized ($\hat x,\hat y$) components generate the standing wave as ${\bold E}_{xy}=2E_1\bigr[\cos(k_0x)\hat y+\cos(k_0y)\hat x\bigr]$. The out-of-plane polarized ($\hat z$) component generates the light field as $E_z\hat z=\hat z E_2(e^{ik_0x}+e^{-ik_0y}+\alpha e^{-ik_0x+i\pi}+\alpha e^{ik_0y+i\pi})$. Here we set that the $\hat z$ component light is partially reflected with a ratio $\alpha$ by the mirror $P$. The $\pi$ phase is the difference of phases acquired by $\hat x$ and $\hat z$ components of the laser beam traveling from the lattice center to mirrors $M_3$ and $P$, respectively, and back to the lattice. From ${\bold E}_{xy}$ and $E_z\hat z$ one can obtain the total lattice potential $V_\text{Latt}(x,y)$ with the amplitudes $\{V_0,V_1,V_2\}\propto \{4|E_1|^2-4\alpha|E_2|^2,4(1+\alpha^2)|E_2|^2,8\alpha|E_2|^2\}$.
Hence the amplitudes $V_{0,1,2}$ can be readily controlled by tuning $\alpha$ and $E_{1,2}$. For example, taking that $\alpha=0.35$ and $|E_1|^2=1.47|E_2|^2$, we have $V_0=8E_R$ and $V_1=8E_R$ if setting $V_2=5E_R$. This regime gives $t_p^X\approx0.1E_R,t_s^Y\approx0.01E_R$, and $\Delta E=E_{p_X}-E_s\approx1.75E_R$, which well meet the former two ingredients of realization.

The configuration for a strong interaction $U_s$ at A sites and weak $U_p$ at B sites can be reached with optical control of Feshbach resonance, which manipulates the atomic interactions by optical field strength~\cite{Walraven1996,Lett2000,Denschlag2004,Takahashi2008,Bauer2009,Ye2011,ChengChin2015}. The staggered modulation of interactions at A and B sites can be tuned with a periodic optical potential which minimizes (maximizes) at A (B) sites. Interestingly, in our case the $V_{1,2}$-terms in the optical lattice potential match the required profile and can play such role. We note that such a spatial modulation of interactions has been achieved in the recent experiment with a long life time~\cite{ChengChin2015}.

The full controllability of cold atoms can enable us to distinguish the CK insulating phase from single-particle QAH states. The present CK insulator can be identified by three characteristic features, namely, a critical $s$-$p_X$ coupling strength $t_{sp}^c$ for CK phase transition, the nontrivial topology in the bulk band, and the Mott behavior of $s$-orbital. The former two features can be confirmed by band topology measurement through Hall effect~\cite{Esslinger2014} which exists for $t_{sp}>t^c_{sp}$ but is absent for $t_{sp}<t^c_{sp}$, while the last one is a key feature which makes the CK phase be exceptional. Since the parent state of $s$-orbital is a Mott insulator, the double occupancy of $s$-orbital is suppressed by the Hubbard gap. This property keeps both before and after the CK phase transition. Therefore, the third feature of CK phase can be detected by measuring the double occupancy of the $s$ and $p_X$ orbitals, which can be carried out in standard cold experiments~\cite{Jordens2008,Takahashi2012}.

\begin{figure}[b]
	\includegraphics[width=0.48\textwidth]{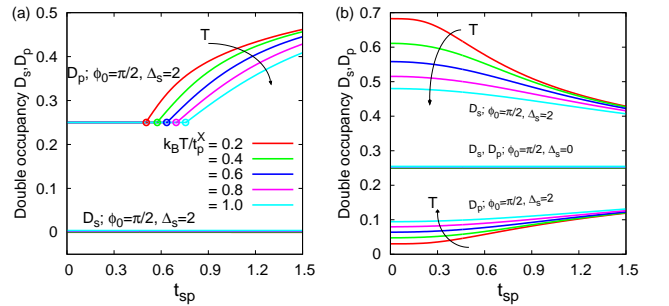}
	\caption{(color online). a) Double occupancy of $p_X$ orbitals $D_p$ and $s$ orbitals $D_s$ with $(t_p^X,t_s^Y,\Delta_s,U_p,U_s)=(1,0.1,2,0,20)$ as a function of $t_{sp}$ (in unit of $t_p^X$) for various sets of temperature $k_\text{B}T$. The circles at $D_p$ curves represent the CK phase transition points. (b) Double occupancy $D_p$ and $D_s$ in single particle QAH phase with $\Delta_s=0$ and $2$.}
	\label{fig:double}
\end{figure}
The double occupancy of $s$-orbital can be derived in the atomic limit. The partition function reads $Z=\sum_{n=0}^2z^n\exp(-\beta E_n)$, where $z=\exp(\beta\mu)$ and $E_n$ is the energy of $n$ atoms occupying the $s$-orbital at an A site. We have that $\{E_0,E_1,E_2\}=\{0,-\Delta_s,U_s-2\Delta_s\}$ and the double occupancy of $s$-orbital $D_s=z^2\exp(-\beta E_2)/Z$. Note that $\mu$ should be determined self-consistently with the total number of atoms in $s$ and $p_X$ states being fixed. On the other hand, the double occupancy of $p_X$-orbital can be obtained by $D_p=(2/N)\partial{\cal F}(U_p,T,t_{sp})/\partial U_p$. From the numerical results shown in Fig~\ref{fig:double} (a), one can find that $D_s$ is greatly suppressed both before and after the CK phase transition, while $D_p$ is unchanged versus $t_{sp}$ in the decoupled phase and increases in the CK phase. The later property reflects that the emergent hybridization in the CK phase leads to a net pumping of atoms from $s$-orbital to $p_X$ orbital. For a comparison, in Fig~\ref{fig:double} (b) we show the results for a single-particle QAH phase. With the same single-particle parameters, the double occupancy $D_s$ is much larger and exhibits qualitatively different behavior upon increasing $t_{sp}$.

In conclusion we have proposed to realize and observe CK insulators in an optical superlattice with laser-assisted $s$ and $p$ orbital hybridization and synthetic gauge field, which can be engineered based on the recently fast developing techniques in cold atoms. The predicted CK insulator can be identified by three characteristic features, namely, the existence of a critical $s$-$p$ coupling strength for CK phase transition, the nontrivial topology in the bulk band, and the Mott behavior of the $s$-orbital. These features distinguish the strongly correlated CK insulating phase from the single-particle QAH states.

\begin{acknowledgments}
We thank Fa Wang for valuable discussions. X.J.L. also thanks Kai Sun, Hui Zhai, and Andriy Nevidomskyy for insightful comments. This work is supported by National Basic Research Program of China (Grants No. 2015CB921102) and National Natural Science Foundation of China (No. 11534001 and No. 11574008). XJL is also supported in part by the Thousand-Young-Talent Program of China.
\end{acknowledgments}

	\widetext
	\section*{Supplementary Material -- Chern Kondo Insulator in an Optical Lattice}
	
\subsection{Double-well optical superlattice}

We provide here more details for the generation of the square superlattice potential $V_\text{Latt}(x,y)$ through the setup shown in Fig.~1 (a) of the main text, with an incident laser field $\bold E=(E_1\hat y+E_2\hat z)e^{ik_0x}$. The initial phases are irrelevant and have been neglected. The light beam is reflected by the mirrors, with the forward and backward traveling fields intersecting and forming lattice in the middle region. In particular, the in-plane polarized ($\hat x,\hat y$) components are reflected by mirrors $M_1$, $M_2$, and $M_3$, and can form the standing wave in the form
\begin{eqnarray}
{\bold E}_{xy}=2E_1e^{i\varphi_1+i\varphi_2}\bigr[\cos(k_0x-\varphi_1-\varphi_2)\hat y+\cos(k_0y+\varphi_2)\hat x\bigr].
\end{eqnarray}
Here $\varphi_1$ represents the phase acquired by the in-plane light beam through the path from lattice center (intersecting point) to the mirror $M_1$, then to $M_2$, and finally back to the lattice center, and $\varphi_2$ represents the phase acquired through the path from lattice center to the mirror $M_3$ [Fig.~1 (a)]. On the other hand, after traveling via $M_1$ and $M_2$, the out-of-plane polarized ($\hat z$) component is reflected by a PBS (polarization beam splitter) and then partially reflected by $P$ with a reflection ratio $\alpha$. By controlling the optical path, we can set that the $\hat z$-component beam acquires an additional $\pi/2$ phase relative to $\varphi_2$ by the in-plane polarized component traveling through the path from lattice center to PBS and then to $P$. This gives the standing wave for the $\hat z$-component light
\begin{eqnarray}
E_z\hat z=\hat z E_2(e^{ik_0x}+e^{-ik_0y+i\varphi_1}+\alpha e^{-ik_0x+i2\varphi_1+i2\varphi_2+i\pi}+\alpha e^{ik_0y+i\varphi_1+i2\varphi_2+i\pi}).
\end{eqnarray}
It is easy to see that the phase factors $\varphi_{1,2}$ can be absorbed by shifting the position $k_0x\rightarrow k_0x+\varphi_1+\varphi_2$ and $k_0y\rightarrow k_0y-\varphi_2$. Then the light fields read
\begin{eqnarray}
{\bold E}_{xy}&=&2E_1\bigr[\cos(k_0x)\hat y+\cos(k_0y)\hat x\bigr],\\
E_z\hat z&=&\hat z E_2(e^{ik_0x}+e^{-ik_0y}+\alpha e^{-ik_0x+i\pi}+\alpha e^{ik_0y+i\pi}).
\end{eqnarray}
The above formulae imply that the generated standing wave fields are stable against phase fluctuations, which at most give rise to global shifts of the lattice and do not affect the model realization. From the above results we can get light field strength that (neglecting the constant terms)
\begin{eqnarray}
|{\bold E}_{xy}|^2&=&4|E_1|^2\bigr[\cos^2(k_0x)+\cos^2(k_0y)\bigr],\\
|E_z|^2&=&4|E_2|^2[(1+\alpha^2)\sin^2[\frac{k_0}{2}(x-y)]+2\alpha\cos^2[\frac{k_0}{2}(x+y)]-4\alpha|E_2|^2\cos^2(k_0x)-4\alpha|E_2|^2\cos^2(k_0y)¡£
\end{eqnarray}

Note that the lattice potential is proportional to the light field strength, with the proportional factor being the dipole matrix element. Therefore, the total lattice potential takes the following form
\begin{eqnarray}
V_\text{Latt}(x,y) = -V_0 [\cos^2(k_0 x) + \cos^2(k_0 y)]-V_1 \sin^2[\frac{k_0}{2}(x-y)]-V_2 \cos^2[\frac{k_0}{2}(x+y)],
\end{eqnarray}
where the amplitudes $\{V_0,V_1,V_2\}\propto \{4|E_1|^2-4\alpha|E_2|^2,4(1+\alpha^2)|E_2|^2,8\alpha|E_2|^2\}$.
Hence the amplitudes $V_{0,1,2}$ can be readily controlled by tuning $\alpha$ and $E_{1,2}$. This is the lattice potential considered in the main text. Taking that $\alpha=0.35$ and $|E_1|^2=1.47|E_2|^2$, we have $V_0=8E_R$ and $V_1=8E_R$ if setting $V_2=5E_R$. This parameter regime gives the lattice potential file shown in Fig.~1(b), with a large anisotropy along $X/Y$ direction. The relevant parameters are obtained as $t_p^X\approx0.1E_R,t_s^Y\approx0.01E_R$, and $\Delta E=E_{p_X}-E_s\approx1.75E_R$. It is noteworthy that the present realization is of topological stability, namely, the phase fluctuations in the incident laser beam cannot affect the configuration of the present double square optical lattice.

\subsection{Slave boson theory}
	
	The local correlation for the mixed $s$ and $p_{X}$ orbital lattice takes usual on-site Hubbard interaction
	\begin{equation}
	\mathcal{H}_\text{I}\equiv U_p\sum_i \hat{n}_{p_Xi\uparrow}\hat{n}_{p_Xi\downarrow}
+U_s\sum_j \hat{n}_{sj\uparrow}\hat{n}_{sj\downarrow}. \nonumber
\end{equation}
When the interaction $U_p$ for $p_X$ orbitals is weak compared with band width of the dispersive $p_X$ orbitals,
the Hartree approximation $ \hat{n}_{p_Xi\uparrow}\hat{n}_{p_Xi\downarrow} \simeq \sum_{\sigma}n_{p_Xi\sigma}\hat{n}_{p_Xi\bar{\sigma}}
-n_{p_Xi\uparrow}n_{p_Xi\downarrow}$ is adequate to capture local correlation effects.
The strong interaction $U_s$ is treated more seriously by introducing auxiliary slave particles to describe the many-body configurations due to the localized nature of $s$ orbitals.
In the slave-boson theory~\cite{Kotliar}, the spin-$\sigma$ fermion operator is written as
\begin{equation}
s_{i\sigma}=\hat{z}_{i\sigma}f_{i\sigma},
\hat{z}_{i\sigma}=\hat{L}_{i\sigma}\left(b_i^\dagger c_{i\sigma}+c_{i\bar{\sigma}}^\dagger d_i\right)\hat{R}_{i\sigma},
\nonumber
\end{equation}
where the boson operators describe the holon $b_i$, singly occupied states $c_{i\sigma}$, doublon $d_i$, and $f_{i\sigma}$ is a fermion operator.
The explicit form of renormalization operators $\hat{L}_\sigma$ and $\hat{R}_\sigma$ takes the form
\begin{eqnarray}
\hat{L}_{i\sigma}&=&\left(1-d_i^\dagger d_i-c_{i\sigma}^\dagger c_{i\sigma}\right)^{-\frac{1}{2}}, \nonumber\\
\hat{R}_{i\sigma}&=&\left(1-b_i^\dagger b_i-c_{i\bar{\sigma}}^\dagger c_{i\bar{\sigma}}\right)^{-\frac{1}{2}}, \nonumber
\end{eqnarray}
recovering the Gutzwiller approximation~\cite{Gutzwiller65} as a saddle point solution of Hubbard model.

To eliminate unphysical states, the Lagrange multipliers $\lambda$ and $\lambda_\sigma$ are then introduced to ensure the local constraint of the completeness of the enlarged Hilbert space
\begin{eqnarray}
\hat{Q}_i\equiv b_i^\dagger b_i+\sum_\sigma c_{i\sigma}^\dagger c_{i\sigma}+d_i^\dagger d_i &=&1£¬ \nonumber
\end{eqnarray}
and two equivalent ways to count the local fermion occupancy for each spin projection $\sigma$ between original fermion and slave boson representations
\begin{eqnarray}
\hat{Q}_{i\sigma}\equiv s_{i\sigma}^\dagger s_{i\sigma}-c_{i\sigma}^\dagger c_{i\sigma}-d^\dagger_{i}d_{i} &=&0. \nonumber
\end{eqnarray}

The partition function within path integral formulism over coherent states of Fermi and Bose fields has the following form
\begin{equation}
\mathcal{Z}\equiv\int \mathcal{D}\left[b,b^\dagger\right] \mathcal{D}\left[d,d^\dagger\right] \mathcal{D}\left[c,c^\dagger\right]
\mathcal{D}\left[f,f^\dagger\right] \mathcal{D}\left[\lambda,\lambda_\sigma\right]
e^{-\int_{0}^{\beta}\mathcal{L}d\tau}, \nonumber
\end{equation}
where $\beta^{-1}\equiv k_\text{B}T$ and the corresponding Lagrangian $\mathcal{L}$ is given by
\begin{eqnarray}
\mathcal{L} &\equiv& \sum_i \left(b_i^\dagger \partial_\tau b_i+d_i^\dagger \partial_\tau d_i\right)
+\sum_{i\sigma}\left(c^\dagger_{i\sigma}\partial_\tau c_{i\sigma}
+f^\dagger_{i\sigma}\partial_\tau f_{i\sigma}\right)
+\sum_i U_p\left(\sum_{\sigma}n_{p_Xi\sigma}\hat{n}_{p_Xi\bar{\sigma}}
-n_{p_Xi\uparrow}n_{p_Xi\downarrow}\right)+\sum_i U_s d_i^\dagger d_i \nonumber	\\
&+&\sum_{i\sigma} [
-t_s^Y z_{i\sigma}^\dagger z_{i\pm\hat{Y}\sigma}f^\dagger_{i\sigma}f_{i\pm\hat{Y}\sigma}-\Delta_s f^\dagger_{i\sigma}f_{i\sigma}
+t_p^X p^\dagger_{Xi\sigma}p_{Xi\pm\hat{X}\sigma}]
+ \sum_{\left<ij\right>\sigma} [F(\mathbf{r})z_{i\sigma}^\dagger f_{i\sigma}^\dagger p_{Xj\sigma}\delta_{j,i+\mathbf{r}}+\text{h.c.}] \nonumber \\
&+&\sum_i\left(\lambda_i \hat{Q}_i+\sum_{\sigma}\lambda_{i\sigma}\hat{Q}_{i\sigma}\right). \nonumber
\end{eqnarray}

In the uniform mean-field slave boson approximation with infinite on-site Coulomb interactions $U_s\to\infty$,
the doublon vanishes to avoid the unphysical divergence and the boson operators are replaced by time- and site-independent numbers $\bar{b}$, $\bar{c}_\sigma$ and $\bar{d}=0$.
The resulting renormalization factor has a concise expansion
\begin{equation}
z_{i\sigma}=\frac{\bar{b} \bar{c}_{\sigma}}
{\sqrt{1-\bar{c}_{\sigma}^2}\sqrt{1-\bar{b}^2-\bar{c}_{\bar{\sigma}}^2}}\equiv Z_\sigma.\nonumber
\end{equation}
The functional for free energy per site is easily obtained from the partition function $\mathcal{F}\equiv-k_\text{B}T\ln\mathcal{Z}/N$ after integration of Fermi fields $\left(f,f^\dagger\right)$ and takes the following form
\begin{equation}
\mathcal{F}=-\frac{1}{N\beta}\sum_{\mathbf{k}\sigma,\pm}\ln \left(1+\exp[-\beta E^{\pm}_{\mathbf{k}\sigma}]\right)
+\lambda\left(\sum_{\sigma}\bar{c}_{\sigma}^2+\bar{b}^2-1\right)
-\sum_{\sigma}\lambda_{\sigma}\bar{c}_{\sigma}^2. \nonumber
\end{equation}
Here,  $E^{\pm}_{{\bf k}\sigma}$ are the lower $(-)$ and upper $(+)$
hybridized bands for the quasi particle with spin projection $\sigma$, respectively:
$E^{\pm}_{{\bf k}\sigma}\equiv\frac{1}{2}\left[
\varepsilon_{p_X\mathbf{k}\sigma}+\varepsilon_{s{\bf k}\sigma}
\pm \sqrt{(\varepsilon_{p_X\mathbf{k}\sigma}-\varepsilon_{s{\bf k}\sigma})^{2}+4|Z_\sigma\mathcal{V}_\mathbf{k}|^2}\right]$,
$\varepsilon_{p_X{\bf k}\sigma}\equiv 2t_p^X\cos k_X
+U_pn_{p_X\bar{\sigma}}$,
$\varepsilon_{s\mathbf{k}\sigma}\equiv -2Z_\sigma^2 t_s^Y\cos k_Y-\Delta_s+\lambda_\sigma$
and $\mathcal{V}_\mathbf{k}\equiv 2it_{sp}\left(\sin\frac{k_X+k_Y}{2}
\exp\left[i\phi_0\right]+\sin\frac{k_X-k_Y}{2}\right)$.
The extremization of the free energy with respect to
the mean field parameters are determined by
$\frac{\partial \mathcal{F}}{\partial \xi}=0$ with $\xi=\bar{c}_\sigma,\bar{b},\lambda_\sigma,\lambda$,
which yields several coupled saddle-point equations
\begin{eqnarray}
\{\lambda-\lambda_{\sigma^\prime},\lambda\}
&=& \sum_{\sigma}\{
\frac{Z_\sigma}{\bar{c}_{\sigma^\prime}}\frac{\partial Z_\sigma}{\partial \bar{c}_{\sigma^\prime}},
\frac{Z_\sigma}{\bar{b}}\frac{\partial Z_\sigma}{\partial \bar{b}}\}
\times
\frac{1}{N}\sum_{{\bf k},\pm}\left[
t_s^Y\cos k_Y\mp \frac{(\varepsilon_{p_X\mathbf{k}\sigma}-\varepsilon_{s\mathbf{k}\sigma})
	t_s^Y\cos k_Y+|\mathcal{V}_\mathbf{k}|^2}
{\sqrt{(\varepsilon_{p_X\mathbf{k}\sigma}-\varepsilon_{s{\bf k}\sigma})^{2}+4|Z_\sigma\mathcal{V}_\mathbf{k}|^2}}\right]
f(E_{{\bf k}\sigma}^{\pm}),	\nonumber\\	
\bar{c}_\sigma^2 &=& \frac{1}{2N}\sum_{{\bf k},\pm}
\left[1\mp\frac{\varepsilon_{p_X\mathbf{k}\sigma}
	-\varepsilon_{s{\bf k}\sigma}}{\sqrt{(\varepsilon_{p_X\mathbf{k}\sigma}
		-\varepsilon_{s{\bf k}\sigma})^{2}+4|Z_\sigma\mathcal{V}_\mathbf{k}|^2}}\right]
f(E^{\pm}_{{\bf k}\sigma}), \nonumber\\	
\bar{b}^2&=& 1-\sum_\sigma \bar{c}_\sigma^2, \nonumber
\end{eqnarray}
where $f(E)$ is the Fermi-Dirac distribution function.

Note that in the main text, the substituted notion $\bar{c}_\sigma\to \bar{s}_\sigma$ and $\left(f_{i\sigma},f^\dagger_{i\sigma}\right)\to \left(s_{i\sigma},s^\dagger_{i\sigma}\right)$,
keeping the formalism complete in essence,
is adopted without expanding the main text to introduce the auxiliary number $\bar{c}_\sigma$ and operator $f_{i\sigma}$.

\subsection{Effective model}

The standard periodic Anderson model is shown to connect the Kondo lattice via the well known Schrieffer-Wolff transformation~\cite{Schrieffer66}.
Below we shall derive the effective Kondo Hamiltonian describing the essential Kondo screening at $t_s^Y=0$ and $U_p=0$ limit. The original Hamiltonian $H=H_p+H_s+H_{sp}$ takes the form
\begin{eqnarray}
H_p&=&\sum_{j\sigma}t^x_pp^\dagger_{Xj\sigma}p_{Xj\pm\hat{X}\sigma} \nonumber\\
H_s&=&-\sum_{i\sigma}\Delta_ss^\dagger_{i\sigma}s_{i\sigma}+U_s\sum_i\hat{n}_{si\uparrow}\hat{n}_{si\downarrow} \nonumber \\
H_{sp}&=&\sum_{\left<ij\right>\sigma}\left[F(\mathbf{r})s^\dagger_{i\sigma}p_{Xj\sigma}\delta_{j,i+\mathbf{r}}+\text{h.c.}\right]. \nonumber
\end{eqnarray}
Atomic projectors operating on $i$-th site $s$ orbital are then introduced as follow
\begin{eqnarray}
\text{Holon subspace}: \mathcal{P}_{i,\text{H}}&\equiv&\left(1-\hat{n}_{si\uparrow}\right)\left(1-\hat{n}_{si\downarrow}\right) \nonumber\\
\text{Single occupied subspace}:
\mathcal{P}_{i,\sigma}&\equiv&\hat{n}_{si\sigma}\left(1-\hat{n}_{si\bar{\sigma}}\right) \nonumber\\
\text{Doublon subspace}:
\mathcal{P}_{i,\text{D}}&\equiv&\hat{n}_{si\uparrow}\hat{n}_{si\downarrow} \nonumber
\end{eqnarray}
Using identity $\mathcal{P}_{i,\text{H}}+\sum_{\sigma}\mathcal{P}_{i,\sigma}+\mathcal{P}_{i,\text{D}}=\mathbf{1}$,
the hybridization between $s$ orbitals and $p_X$ orbitals can be written in projected subspace
\begin{eqnarray}
H_{sp}=\sum_\mathbf{k}\left[\mathcal{H}_1(\mathbf{k})+\mathcal{H}_2(\mathbf{k})+\text{h.c.}\right] \nonumber
\end{eqnarray}
with
\begin{eqnarray}
\mathcal{H}_1(\mathbf{k}) \equiv \underset{i\sigma}{\sum}\mathcal{V}\left(\mathbf{k}\right)\mathcal{P}_{i,\sigma}s_{i\sigma}^{\dagger}\mathcal{P}_{i,\text{H}}\frac{1}{\sqrt{N}}p_{X\mathbf{k}\sigma}\exp[-i\mathbf{k}\cdot i], \nonumber\\
\mathcal{H}_2(\mathbf{k}) \equiv
\underset{i\sigma}{\sum}\mathcal{V}\left(\mathbf{k}\right)\mathcal{P}_{i,\text{D}}s_{i\sigma}^{\dagger}\mathcal{P}_{i,\bar{\sigma}}\frac{1}{\sqrt{N}}p_{X\mathbf{k}\sigma}\exp[-i\mathbf{k}\cdot i]. \nonumber
\end{eqnarray}
A little algebra yields several commutation relations $\left[\mathcal{H}_1(\mathbf{k}),H_s\right]=\Delta_s \mathcal{H}_1(\mathbf{k}),
\left[\mathcal{H}_2(\mathbf{k}),H_s\right]=(\Delta_s-U_s) \mathcal{H}_2(\mathbf{k})$.
For the localized $s$ orbitals embedded in a conduction $p_X$ band, the present model can be mapped into a spin exchange Hamiltonian through the explicit canonical transformation $e^{-\mathcal{S}}$~\cite{Schrieffer66}.
The effective Hamiltonian up to quadratic order in $\mathcal{V}^2\left(\mathbf{k}\right)$ has a perturbation expansion via the Baker-Campbell-Hausforff formula and takes the form
\begin{eqnarray}
H_{\text{eff}}=e^{\mathcal{S}}He^{-\mathcal{S}}=H+\left[\mathcal{S},H\right]+\frac{1}{2!}\left[\mathcal{S},\left[\mathcal{S},H\right]\right]+\mathcal{O}\left(\mathcal{V}^3\left(\mathbf{k}\right)\right) \nonumber
\end{eqnarray}
At the first order in $\mathcal{V}_\mathbf{k}$, the Schrieffer-Wolff transformation is constructed to eliminate the hybridization term with
\begin{equation}
\mathcal{S}= \underset{\mathbf{k}}{\sum}
\left[-\frac{1}{\Delta_{s}}\mathcal{H}_{1}\left(\mathbf{k}\right)
-\frac{1}{\Delta_{s}-U}\mathcal{H}_{2}\left(\mathbf{k}\right)-\text{h.c.}\right]. \nonumber
\end{equation}
By collecting the next leading order (quadratic order),
we end up with the exchange Hamiltonian at large $U_s$ limit
\begin{eqnarray}
H_\text{ex} = \frac{1}{4N}\underset{\mathbf{kk^{\prime}}}{\sum}\left(\frac{1}{\Delta_{s}}-\frac{1}{\Delta_{s}-U_s}\right)
\{\underset{i}{\sum}\mathcal{V}\left(\mathbf{k}\right)\mathcal{V}^{*}\left(\mathbf{k}^{\prime}\right)\exp[i\mathbf{k}^{\prime}\cdot i-i\mathbf{k}\cdot i]
p_{X\mathbf{k}^{\prime}}^{\dagger}\vec{\sigma}p_{X\mathbf{k}}s_{i}^{\dagger}\vec{\sigma}s_{i}+\mathbf{k}\longleftrightarrow\mathbf{k}^{\prime}\} \nonumber
\end{eqnarray}
Fourier transforming back to real space, we arrive at a concise exchange Hamiltonian
\begin{eqnarray}
H_\text{ex} = J \underset{\left\langle ij\right\rangle}{\sum}\underset{\left\langle ij^{\prime}\right\rangle}{\sum}
F^{*}\left(\mathbf{r}\right)\delta_{j,i+\mathbf{r}}
F\left(\mathbf{r}^{\prime}\right)\delta_{j^{\prime},i+\mathbf{r}^{\prime}}
p_{Xj}^{\dagger}\vec{\sigma}p_{Xj^\prime}s_{i}^{\dagger}\vec{\sigma}s_{i}\nonumber
\end{eqnarray}
where $J\approx \frac{U_s}{2\Delta_s(U_s-\Delta_s)}$. To summarize, the minimal model $H_\text{eff}=H_p+H_\text{ex}+\cdots$ captures the essential physics.
The exchange Hamiltonian $\mathcal{H}_\text{ex}$ describes that the itinerant $p_X$ bands hybridize with localized $s$ orbitals, giving rise to the effective Kondo coupling.
The screening of localized spin states of $s$ orbitals we studied in the main text is adiabatically connecting the $t_s^Y=0$ and weak $U_p$ limit.
On the other hand, we remark that, in the main text, the Kondo screening is robust against the perturbative interaction $U_p$ which can be tuned by optical control of Feshbach resonance and then be treated by Hartree approximation.
The essential cotunneling process is explicitly captured by the exchange Hamiltonian $\mathcal{H}_\text{ex}$.

\subsection{Double occupancy in atomic limit}

At the half-filling condition, to create a double occupancy  in the lattice costs the energy $U$ for the repulsive Hubbard interaction.
Therefore, the strong repulsive interaction $U$ can greatly suppress the double occupation of lattice sites and lead to Mott insulating phase.
For the fermions in an optical lattices, the techniques to measure the fraction of double occupied sites have been well developed and demonstrated to detect the Mott insulating phase successfully~\cite{Jordens2008,Takahashi2012}.
With a large repulsion $U_s$, the hopping process for $s$ orbital is greatly suppressed. In this case, we consider the atomic limit to approximately calculate the double occupancy by neglecting the kinetic energy. The partition function for $i$-th lattice site via Boltzmann factor has the following form
\begin{equation}
Z_i=\sum_n z^n\exp\left[-\beta E_{i,n}\right],\nonumber
\end{equation}
where $z=\exp\left[\beta\mu\right]$ is the fugacity, $\mu$ is the chemical potential, $n\in\{0,1,2\}$ labels three possible occupation numbers of fermions and $E_{i,n}$ is the corresponding energy.
We take the uniform approximation and drop the lattice site dependence.
The local energy level for $s$ orbitals is $E_{n=\{0,1,2\}}=\{0,-\Delta_s,-2\Delta_s+U_s\}$.
The chemical potential is located by the occupation number $n_s$ in $s$ orbitals
\begin{equation}
n_{s}=\frac{2z\exp\left(\beta\Delta_{s}\right)+2z^{2}\exp\left(2\beta\Delta_{s}-\beta U_s\right)}
{1+2z\exp\left(\beta\Delta_{s}\right)+z^{2}\exp\left(2\beta\Delta_{s}-\beta U_s\right)}. \nonumber
\end{equation}
Before the Chern Kondo phase transition, the chemical potential $\mu$ is fixed at half-filling for $s$-band itself and the particle number $n_s=1$. On the other hand, after the Chern Kondo phase transition, $\mu$ should be determined by the self-consistent solutions. The double occupancy reads
\begin{equation}
D_s = \frac{\bar{z}^2\exp\left[-\beta U_s\right]}
{1+2\bar{z}+\bar{z}^2\exp\left[-\beta U_s\right]}\nonumber
\end{equation}
where $\bar{z}=\frac{\sqrt{1+\frac{4n_s}{2-n_s}}\exp\left[-\beta U_s\right]-1}{2\exp\left[-\beta U_s\right]}$. At high temperatures, the double occupancy saturates at $D_s=\frac{\left(\sqrt{2+3n_s}-\sqrt{2-n_s}\right)^4}{16n_s^2}$.
While, the asymptotic expansion at low temperature limit takes the form $D_s=\frac{n_s^2}{4-n_s^2}\exp\left[-\beta U_s\right]$ showing the exponential behavior with respect to inverse temperature $\beta$ and Hubbard interaction $U_s$.
The double occupancy is negligible small at large $U_s$ relative to temperature $k_\text{B}T$, which is consistent with the vanishing doublon assumption in slave boson calculations.

\end{document}